\documentclass[seceq]{ptptex}

\usepackage
{graphicx}

\newcommand{\del}{\partial}

\newcommand{\e}{\epsilon}
\newcommand{\ve}{\varepsilon}
\newcommand{\sgn}{{\rm sgn}}

\title{
Non-Abelian Gauge Field Localized on \\
Walls with Four-Dimensional World Volume
}

\author{
Kazutoshi \textsc{Ohta}$^{1,}$\footnote{E-mail: kohta@law.meijigakuin.ac.jp} 
and Norisuke \textsc{Sakai}$^{2,}$\footnote{E-mail: sakai@lab.twcu.ac.jp}
}

\inst{
$^1$Institute of Physics, Meiji Gakuin University, 
Yokohama 244-8539, Japan\\
$^2$Department of Mathematics, Tokyo Woman's Christian University, 
Tokyo 167-8585, Japan
}

\abst{
A mechanism using the position-dependent gauge coupling 
is proposed to localize non-Abelian gauge 
fields on domain walls in five-dimensional space-time. 
Low-energy effective theory posseses a massless vector 
field, and a mass gap. 
The four-dimensional gauge invariance is maintained intact. 
We obtain perturbatively the four-dimensional Coulomb law 
for static sources on the domain wall. 
BPS domain wall solutions with the localization mechanism 
are explicitly constructed in the $U(1)\times U(1)$ 
supersymmetric gauge theory coupling to the non-Abelian 
gauge fields only through the 
cubic prepotential, which is 
consistent with the general principle of supersymmetry in 
five-dimensional space-time. 
}

\begin{document}

\maketitle

\section{
Introduction}\label{sc:introduction}

Besides supersymmetric theories, \cite{Dimopoulos:1981zb} 
the most intriguing possibilities for unified theories 
beyond the standard model are models with extra dimensions, 
that is, the brane-world scenario 
\cite{Horava:1995qa,ArkaniHamed:1998rs,Randall:1999ee}. 
In this scenario, our four-dimensional 
world is localized on defects (which are often called 
branes) such as domain walls in a higher-dimensional 
space-time, and the standard model particles are assumed 
to be localized on the defect. 
Domain walls are the simplest defect and have been most 
useful to construct realistic models. 
However, localization of gauge fields on domain walls 
has been notoriously difficult in field theories, although 
scalar and spinor fields have been successfully localized. 
It has been recognized early that the model with the warp 
factor does not help to localize the gauge field on the domain wall. 
\cite{Davoudiasl:1999tf,Pomarol:1999ad} 
More recently it was proposed that 
bulk and boundary mass terms 
can be introduced into a warped model to help 
localize gauge fields.
\cite{Batell:2005wa} 
It was achieved at the cost of a subtle 
fine-tuning as well as certain boundary interactions 
to restore the gauge invariance. 
An explicit model of Abelian gauge field localized on 
a domain wall in five-dimensional space-time has been 
obtained using tensor multiplet \cite{Isozumi:2003uh}. 
However, it has been difficult to extend the idea of 
tensor multiplet to incorporate the non-Abelian local gauge 
symmetry. 
If we are content with toy models of lower-dimensional 
world-volume, 
such as domain walls with the three-dimensional world-volume, 
there have been a number of proposals, assuming nonperturbative effects 
\cite{Dvali:1996xe}, or using perturbative 
methods \cite{Shifman:2002jm,Dubovsky:2001pe}. 
However, it has been difficult to obtain an explicit model 
of non-Abelian gauge fields localized on a domain wall 
in five-dimensional space-time.

A basic problem has been pointed out to 
localize gauge fields on a domain wall
\cite{Dvali:1996xe,ArkaniHamed:1998rs}. 
We wish to obtain a (perturbatively) unbroken gauge symmetry 
on the world volume of the domain wall. 
Since one wishes to prevent gauge fields to propagate 
freely in the bulk, one is tempted to consider that 
the bulk space-time outside of the domain wall to be in 
the Higgs phase where gauge symmetry is broken. 
Unfortunately, however, the flux coming out of the source 
on the domain wall is absorbed by the bulk in the Higgs phase 
and cannot reach beyond the width of the domain wall 
even in the direction along the world volume of the domain wall. 
Because of this screening effect, the vector field 
acquires a mass of the order of the inverse width of the 
wall \cite{Dvali:1996xe,ArkaniHamed:1998rs,Isozumi:2003uh,Maru:2003mx}. 
On the contrary, if a vector field is confined in the bulk 
and deconfined on the domain wall, the flux coming out of a source 
should be expelled from the bulk, producing the four-dimensional 
Coulomb law on the world volume of the domain wall. 
Therefore we need to consider the confining phase for 
the bulk, whose explicit implementation is often difficult. 
This difficulty is particularly acute in our problem, since 
a realistic model requires the five-dimensional bulk, where 
the knowledge on nonperturbative effects is scarce. 
Nonperturbative effects in a five-dimensional gauge theory 
with a cut-off was proposed to obtain the layered phase 
that confines only along the extra dimension\cite{Fu:1983ei}. 

Even if the detailed knowledge of nonperturbative effects 
is not available, the confining medium can be 
rephrased classically by introducing a dielectric 
permeability $\epsilon$ for gauge fields\cite{Kogut:1974sn,
Fukuda:1977wj}. 
In the classical electrodynamics of a dielectric medium, 
the electric flux density $\boldsymbol{D}$ is the sum of 
the electric field $\boldsymbol{E}$ and the polarization 
$\boldsymbol{P}$ induced by $\boldsymbol{E}$ 
\begin{equation}
\boldsymbol{D}=\epsilon_0 \boldsymbol{E}+\boldsymbol{P}
\equiv \epsilon \boldsymbol{E}
\end{equation}
where the permeability $\epsilon(\boldsymbol{x})$ depends on the 
spatial position $\boldsymbol{x}$, and the 
vacuum permeability is denoted as $\epsilon_0$. 
For ordinary dielectric media, the dielectric permeability 
$\epsilon$ is greater than the vacuum permeability 
$\epsilon_0$, since the polarization $\boldsymbol{P}$ 
is always induced in the same direction as the electric 
field $\boldsymbol{E}$. 
On the other hand, it has been proposed that the confining 
vacuum can be represented by an unusual dielectric permeability 
$\epsilon \to 0$, namely by a perfect dia-electric medium. 
A relativistic version of the dielectric permeability can be 
expressed by a Lagrangian \cite{Kogut:1974sn} with the 
field strength $F_{\mu\nu}\equiv \partial_\mu W_\nu-\partial_\nu W_\mu$
of an Abelian vector field $W_\mu$ 
\begin{equation}
{\cal L}=-{1\over 4} \epsilon(\boldsymbol{x}) F_{\mu\nu}F^{\mu\nu} . 
\end{equation}
The dielectric permeability in this form is nothing but the 
position-dependent gauge coupling $g^2(\boldsymbol{x})$ 
\begin{equation}
\epsilon(\boldsymbol{x}) = {1 \over g^2(\boldsymbol{x})} ,
\end{equation}
and the region of the confining vacuum is represented 
by the strong coupling $g^2 \to \infty$ in this classical 
language. 
Therefore we can represent the confining vacuum in the bulk 
and deconfining vacuum on the domain wall classically by the 
position-dependent gauge coupling, by 
requiring strong coupling $g^2 \to \infty$ asymptotically 
in the bulk away from the domain wall, and weak coupling 
on the domain wall.

The purpose of our paper is to propose 
the position-dependent gauge coupling 
as a model for a gauge field localization on domain walls, 
to examine its generic properties, and to construct explicit 
examples of such domain walls as BPS states in 
supersymmetric gauge theories. 
The general features of the position-dependent gauge coupling are discussed in \S2.
We find that the four-dimensional gauge invariance 
is intact, assuring the existence of massless gauge field 
in the low-energy effective theory. 
By analyzing modes of effective fields 
in four-dimensional world-volume, we find 
that our model indeed has a massless 
gauge field and a mass gap. 
It is worth stressing that this mechanism is applicable to 
localization of the non-Abelian gauge fields as well as the 
Abelian gauge field.
We will discuss these points in \S3.
We also show in \S4 that the 
gauge field localized on the domain wall exhibits
perturbatively a four-dimensional Coulomb law 
for static sources localized on the world volume of 
the domain wall. 
As explicit examples, we construct BPS domain wall solutions 
in supersymmetric gauge theories with $U(1)\times U(1)$ 
gauge group in \S5. 
The non-Abelian gauge fields couple to these 
$U(1)\times U(1)$ vector multiplets only through the cubic 
prepotential and are localized on the domain wall. 
It is remarkable that the cubic coupling among 
vector multiplets is just sufficient 
to give a nontrivial profile of position-dependent 
coupling function automatically, once the domain wall is 
formed as the background solution. 
This satisfies the stringent constraint of 
supersymmetric gauge theories 
in five-dimensional space-time allowing only up to cubic 
coupling in the prepotential of vector multiplets 
\cite{Seiberg:1996bd}. 
We will discuss briefly mechanisms to introduce 
matter multiplets in the nontrivial representations of 
the localized non-Abelian gauge fields in \S6.

\section{A model of position-dependent gauge coupling
}\label{sc:position_dep_coupl}
In this section, we will explore general features of 
localized gauge fields on domain walls, without relying on 
a particular form of domain wall solutions. 
Concrete examples of domain wall solutions demonstrating 
the feasibility of our mechanism of localization will 
follow in later sections. 
We consider a domain wall in $4+1$ dimensional space-time, 
whose coordinates are denoted by $x^M, M=0, 1, \cdots, 4$, 
whereas the world-volume coordinates and the codimension 
of the domain wall are denoted as $x^\mu, \mu=0,1,2,3$, and 
$y$, respectively. 
Namely the domain wall profile depends only on $y$. 
We assume that the five-dimensional gauge field 
acquires a position-dependent gauge coupling function 
$\e(y)$, on the background domain wall solution. 
We normalize the position-dependent gauge coupling function 
$\e(y)$ as 
\begin{equation}
\int_{-\infty}^\infty dy \,\e(y) =\frac{1}{g_4^2} , 
\label{eq:profile_function}
\end{equation}
and denote the four-dimensional gauge coupling as $g_4$. 
Let us first consider gauge field $W_M^a$ with non-Abelian gauge group $G$,
where $a$ runs from 1 to $\dim G$, in 
$4+1$-dimensional space-time. 
Denoting the source and the field strength as $J_M$ 
and $F^a_{MN}\equiv \partial_M W_N^a - \partial_N W^a_M-f^{abc}W^b_MW^c_N$ 
with $M, N =0, 1, \cdots, 4$, 
we take the following Lagrangian with the position-dependent 
gauge coupling function $\e(y)$ instead of the ordinary 
five-dimensional gauge coupling constant $1/g_5^2$ 
\begin{equation}
{\cal L}=-\frac{1}{4}\e(y)F^{MNa}F^{a}_{MN}.
\label{eq:gauge_field_lagrangian}
\end{equation}
Our metric convention is $\eta_{MN}={\rm diag.}(+, -, \cdots, -)$.

We assume that the position-dependent gauge coupling function 
$\e(y)$ is real and nonnegative everywhere 
\begin{equation}
\e(y) > 0 .
\end{equation}
and 
has a profile localized on a domain wall. 
Let us take the center of the domain wall to be at $y=0$. 
We make a crucial assumption for $\e(y)$ 
to vanish at both infinity $y\to \pm \infty$ 
\begin{equation}
\e(y) \to 0, \qquad y \to \pm \infty . 
\end{equation} 
This asymptotic behavior represents the strong coupling in the bulk, 
namely the confining vacuum. 
It has been proposed that this type of ``a perfect dia-electric medium'' is the classical representation of confining vacuum 
\cite{Kogut:1974sn,Fukuda:1977wj}.

The Lagrangian (\ref{eq:gauge_field_lagrangian}) gives the 
field equation 
\begin{equation}
0 = \epsilon(y) (D^N F_{NM})^a + (\del^y \e(y))F^a_{yM},
\label{non-linear eom}
\end{equation}
where $(D^N F_{NM})^a\equiv \del^NF^a_{NM}-f^{abc}W^{Nb}F^c_{NM}$ 
and 
we have used the fact that the position-dependent gauge coupling 
function depends only on $y$. 
By analyzing the field equation (\ref{non-linear eom}), 
we will explore the low-energy effective field theory in 
four-dimensional space-time in the following sections.

\section{Mode Analysis
}\label{sc:mode_analysis}

In this section, we would like to find massless and massive 
modes that appear in the low-energy effective field theory of the 
five-dimensional gauge theory with the position-dependent 
gauge coupling function $\epsilon(y)$. 
We first need to obtain mode functions using the linearized 
field equation arising from quadratic terms of the 
Lagrangian (\ref{eq:gauge_field_lagrangian}). 
We can define a mode expansion of the Lagrangian assuming 
these mode functions are complete, provided they are 
normalizable. 
The linearized field equation is given by 
\begin{equation}
0=\e(y)(\del^N\del_N W^a_M-\del^N\del_M W^a_N)
+(\del^y\e(y))(\del_yW^a_M-\del_MW^a_y).
\label{eq:linearized_eq_motion}
\end{equation}
To find mode functions, we here choose an axial gauge 
\begin{equation}
W^a_y=0.
\label{eq:axial_gauge}
\end{equation}
Then the field equation 
(\ref{eq:linearized_eq_motion}) for $M=\mu$ component becomes 
\begin{equation}
0=\e(y)\left\{
(\del^\nu\del_\nu-\del_y^2)W^a_\mu - \del_\mu\del^\nu W^a_\nu
\right\}
-
\del_y \e(y)\del_y W^a_\mu.
\label{EOM mu component}
\end{equation}
Operating by $\del^\mu$ to the above field equation, we find
\[
0= -\del_y\left(
\e(y)\del_y(\del^\mu W^a_\mu)
\right).
\]
Thus $\del^\mu W^a_\mu$ separates variables $x$ and $y$, 
and factorizes into
\begin{equation}
\del^\mu W^a_\mu= Y(y)F^a(x),
\label{YF}
\end{equation}
where we define
\begin{equation}
Y(y) = \int^y dy' \frac{1}{\e(y')},
\label{Y of y}
\end{equation}
and $F^a(x)$ is a function of $x$ only. 
Therefore we find no propagating modes in the longitudinal part 
 $\del^\mu W^a_\mu$. 
Even though the longitudinal part does not provide a 
propagating modes, it may contribute to a potential 
responding to a static source, as is usual in quantum 
electrodynamics. We will discuss the static potential 
due to a source in the next section.

Now we decompose $W^a_\mu$ into transverse and longitudinal 
components, $W_\mu^{aT}$ and $W_\mu^{aL}$. 
In particular, using (\ref{YF}), the longitudinal component 
reduce to
\begin{eqnarray}
W_\mu^{aL} &\equiv & \frac{1}{\del^2}\del_\mu \del^\nu W^a_\nu
\nonumber\\
&=& Y(y)\frac{1}{\del^2}\del_\mu F^a(x)
\label{longitudinal W}
\end{eqnarray}
Plugging the decomposition 
$W^a_\mu=W_\mu^{aT}+W_\mu^{aL}$ into (\ref{EOM mu component})
and using (\ref{longitudinal W}), we finally find the 
field equation for the transverse component $W_\mu^{aT}$
\begin{equation}
0=\e(y)\del^\nu\del_\nu W_\mu^{aT}-\del_y(\e(y)\del_yW_\mu^{aT}).
\label{eom}
\end{equation}

Let us now expand $W_\mu^{aT}$ by a complete set of wave 
functions $u_n(y)$ in $y$ as 
\begin{equation}
W_\mu^{aT} = \sum_n w^{a(n)}_\mu(x)u_n(y),
\label{mode expansion}
\end{equation}
where  $w^{a(n)}_\mu(x)$ satisfies
\[
(\del^\nu\del_\nu + m_{n}^2) w^{a(n)}_\mu(x)=0,
\]
and $m_n$ represents mass of the mode. 
Plugging the mode expansion (\ref{mode expansion}) into 
the field equation (\ref{eom}), 
we find the equation for the mode function $u_n(y)$ 
\begin{equation}
\left[
-\frac{1}{\e(y)}\frac{d}{dy}\e(y)\frac{d}{dy}-m_{n}^2
\right]
u_n(y)=0.
\label{wave equation}
\end{equation}
If we change the variable from $y$ to $Y$ 
defined in Eq.~(\ref{Y of y}) satisfying 
$\frac{d}{dY} \equiv \e(y)\frac{d}{dy}$, 
the mode equation (\ref{wave equation}) can be transformed 
into a bound state problem at the threshold (zero energy) 
for a potential $U(Y)$ which is proportional to the mass 
squared $m_{n}^2$ of the mode 
\begin{equation}
H\equiv 
-\frac{1}{2}\frac{d^2}{dY^2}+U(Y)
, \quad 
U(Y) = -\frac{1}{2}m_{n}^2\e^2(y(Y)),
\label{potential}
\end{equation}
\begin{equation}
Hu_n(Y)=0
.
\label{wave equation2}
\end{equation}
Let us note that the normalization of the position-dependent 
coupling function $\epsilon$ is fixed by the 
four-dimensional gauge coupling $g_4$ in 
Eq.~(\ref{eq:profile_function}) 
\begin{equation}
\frac{1}{g_4^2} 
=\int dy \,\e(y) 
=\int dY \,\e^2(y(Y)),  
\label{eq:4dcoupling}
\end{equation}
where the integration should be carried out over the entire 
region of $y$ or $Y$. 

For a given shape $\epsilon(y(Y))$ of the potential, 
we can find all possible threshold bound states 
at various discrete depths of the potential $U(Y)$ 
by adjusting $m_{n}^2$. 
In this way, finding threshold (zero energy) bound state 
solutions gives a discrete spectrum of mass squared $m_{n}^2$ 
of modes. 
We wish to obtain the low-energy effective Lagrangian 
defined by an integration of the fundamental five-dimensional 
Lagrangian over $y$ 
\begin{equation}
{\cal L}_{\rm eff}\equiv \int dy \, {\cal L}_5 .
\label{eq:effective_lag}
\end{equation}
This effective Lagrangian dictates the normalization 
condition which naturally contains the following measure 
$g_4^2\; \epsilon(y)$ for the mode function 
\begin{equation}
\int_{-\infty}^\infty dy \, [g_4^2\; \epsilon(y)]u_n(y)^* u_l(y) 
=\delta_{nl}. 
\label{eq:measure}
\end{equation}
This measure is a distinctive feature of our threshold 
bound state mode functions.

We see that constant mode is always a 
zero energy solution: 
\begin{equation}
u_0={\rm constant}, \quad Hu_0=0. 
\label{eq:constant_mode}
\end{equation}
The normalizability of 
this constant mode is equivalent to the condition of 
finiteness of the effective four-dimensional gauge 
coupling in Eq.~(\ref{eq:4dcoupling}). 
As will be illustrated by solvable examples in the following, 
there is a mass gap for these 
threshold bound states, and their wave functions are 
normalizable because of the nontrivial measure in 
Eq.~(\ref{eq:measure}), 
similarly to the constant mode, provided the finite 
four-dimensional gauge coupling can be defined by 
Eq.~(\ref{eq:4dcoupling}). 
This situation is quite different from the 
threshold bound states in the usual quantum mechanics 
problems in one spatial dimension.

To illustrate our procedure of finding the spectrum of 
modes for the position-dependent gauge coupling, 
we will take two examples of solvable potentials that 
can also serve as approximations to our concrete examples 
of domain wall solutions in subsequent sections.

\bigskip
\hspace*{-\parindent}\underline{Solvable Example 1}

\smallskip

\begin{figure}[t]
\begin{center}
\begin{tabular}{cc}
\includegraphics[scale=0.3]{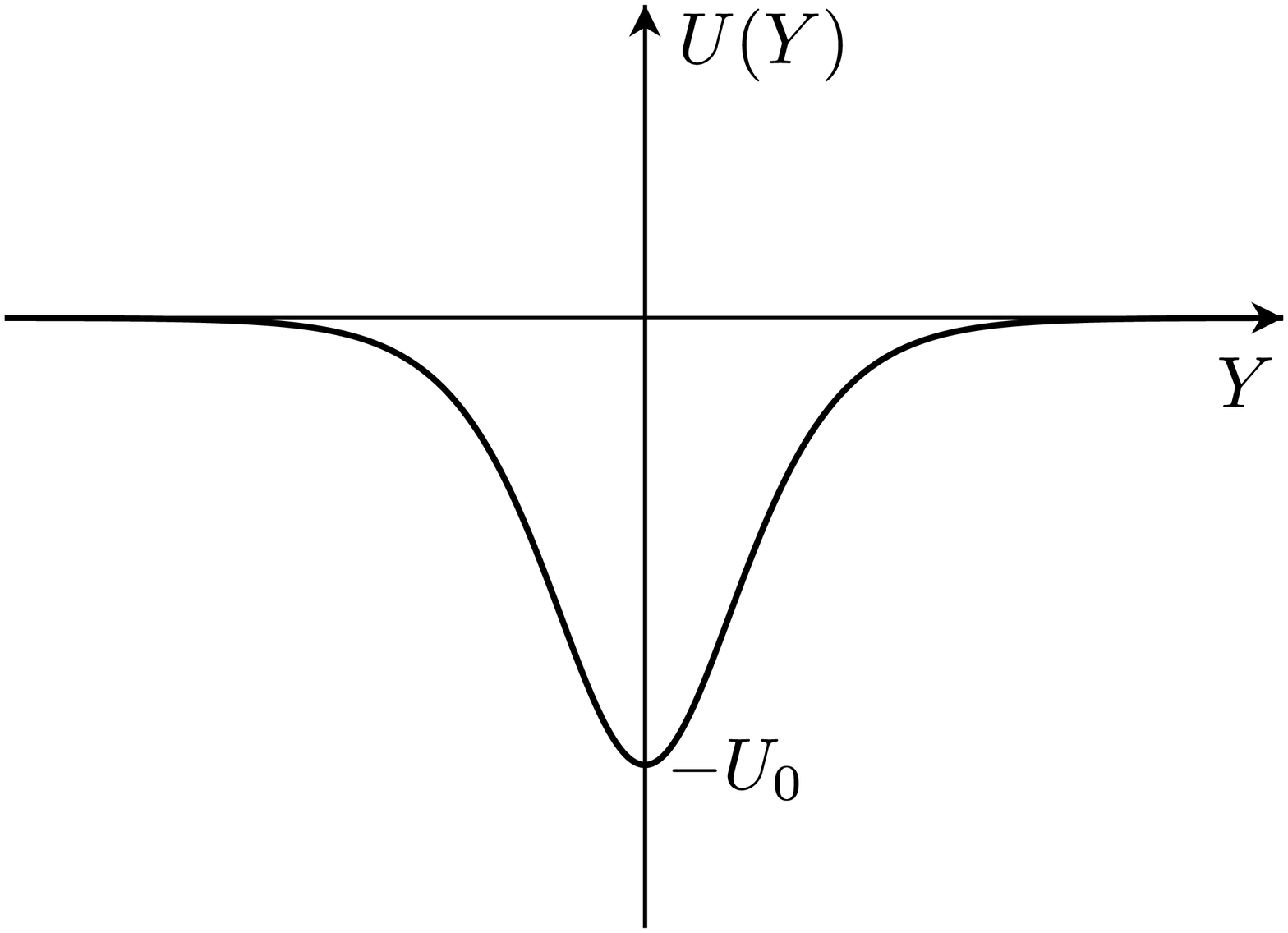} &
\includegraphics[scale=0.37]{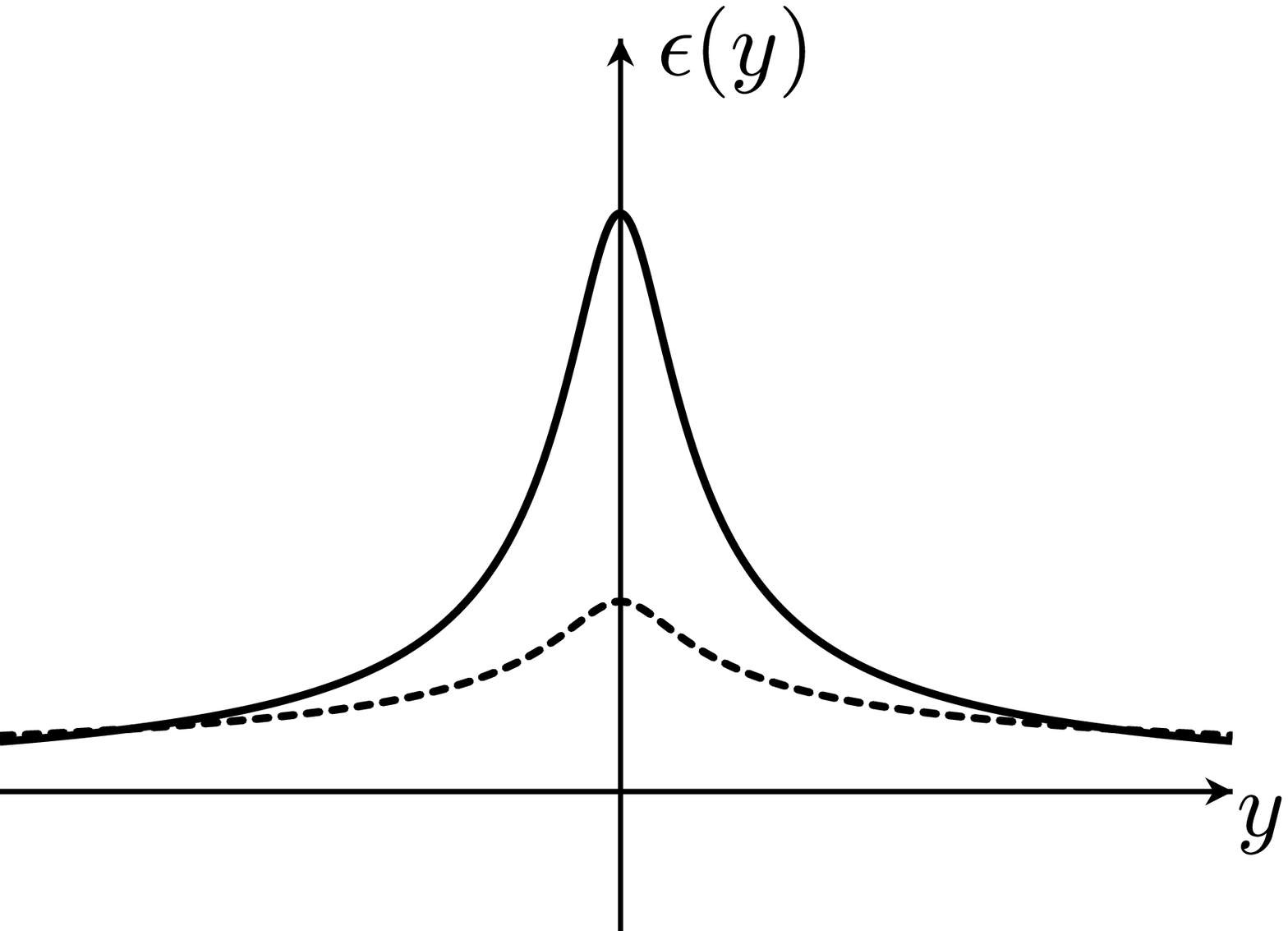}\\
(a) Potential in $Y$ & (b) Position-dependent gauge coupling in $y$
\end{tabular}
\end{center}
\caption{The cosh-type potential and related profile of the coupling function. A dashed line within
the coupling
function represents the wave function of the localized massless vector field around the domain wall.}
\label{cosh-potential plot}
\end{figure}

We first consider the potential
\begin{equation}
U(Y) = -\frac{U_0}{\cosh^2 \alpha Y},
\label{cosh potential}
\end{equation}
which is plotted in Fig.\ref{cosh-potential plot}(a). 
Because of the relation (\ref{potential}) and 
the normalization condition (\ref{eq:4dcoupling}), 
the position-dependent coupling function 
$\epsilon$ is fixed in this case as 
\begin{equation}
\e(Y) =  \frac{1}{g_4}\sqrt{\frac{\alpha}{2}}\frac{1}{\cosh \alpha Y}.
\label{eq:coupling_function1}
\end{equation}
Then Eq.~(\ref{Y of y}) implies 
$y=(g_4\sqrt{2}/\alpha^{3/2})\sinh \alpha Y$.
Therefore, from Eqs.(\ref{potential}) and 
(\ref{cosh potential}), we find
\begin{eqnarray}
\e(y) &=& \frac{1}{g_4}\sqrt{\frac{\alpha}{4}}\frac{1}{\cosh \alpha Y(y)}
=\frac{1}{g_4}\sqrt{\frac{\alpha}{4g_4^2+2g_4\alpha^3 y^2}}, 
\end{eqnarray}
which is shown in Fig.\ref{cosh-potential plot}(b).

If we consider the eigenvalue problem $Hv_k=E_kv_k$, 
we immediately find\cite{Landau} that the finite number of 
bound states exist with a discrete energy spectrum 
$E_k=-(\alpha^2/2)(s-k)^2$, $k=0,1,\ldots \le s$ 
with 
\begin{equation}
s \equiv 
{1\over 2}\left(-1+\sqrt{1+\frac{8U_0}{\alpha^2}}\right)
\label{eq:def_s}
\end{equation}
The threshold bound state $E_k=0$ occurs if and only if 
$k=s$ with $s$ being a nonnegative integer $s=n, n=0,1,2,\ldots$. 
This gives the $n$-th threshold bound state. 
In that case, the potential depth $U_0$ satisfies 
\begin{equation}
U_0 = \frac{\alpha^2}{2}n(n+1) .
\label{eq:potential_height}
\end{equation}
Eq.(\ref{potential}) implies the threshold bound state 
spectrum\footnote{
The continuum spectra with positive energy $E_k$ do not 
contribute, except the limiting case of zero energy $E_k\to 0$. 
By regularizing in a finite interval in $Y$, we find that 
the zero energy solution reduces to our $n=0$ solution in the 
limit of infinite interval. 
}
\begin{eqnarray}
m_n^2 &=& 2g_4^2\alpha n(n+1), \qquad n=0,1,2,\ldots. 
\end{eqnarray}

First we consider the energy level at $n=0$. 
This gives the massless mode $m_0^2=0$, whose wave function 
turns out to be the constant $u_0(y)=1$, as we have seen before. 
Secondly, the first excited mode 
gives $m_1^2 = 4 g_4^2 \alpha$. 
Similarly, higher excited modes gives larger discrete 
values of $m_n^2$. 
The normalizability (\ref{eq:measure}) of the $n$-th mode 
in this example is given by the finiteness of 
\begin{equation}
\int_{-\infty}^\infty dy \, \e(y)|u_n(y)|^2
= \int_{-\infty}^\infty dY \, \frac{\alpha}{2g_4^2 \cosh^2 \alpha Y}|u_n(Y)|^2.
\end{equation}
We find that wave functions of all the excited modes are 
normalizable, since they are just polynomials\cite{Landau} 
in $\tanh \alpha Y$. 
Thus we conclude that there is a massless mode 
and a mass gap for the first excited mode, both of 
which are  normalizable. 
We can safely use the effective field theory of the 
massless gauge fields below the energy scale $m_1$, 
ignoring the massive modes. 
The value of the mass gap is proportional to the
inverse of the width $\frac{1}{\alpha}$ of the position-dependent 
coupling function $\epsilon$ and the square of the gauge coupling 
$g_4^2$. 
The massless zero mode wave function including the 
square root of the measure $\sqrt{\epsilon(y)}u_0(y)$ is plotted 
in Fig.\ref{cosh-potential plot}(b) by a dashed line.

\bigskip
\hspace*{-\parindent}\underline{Solvable Example 2}

\smallskip

\begin{figure}[t]
\begin{center}
\includegraphics[scale=0.3]{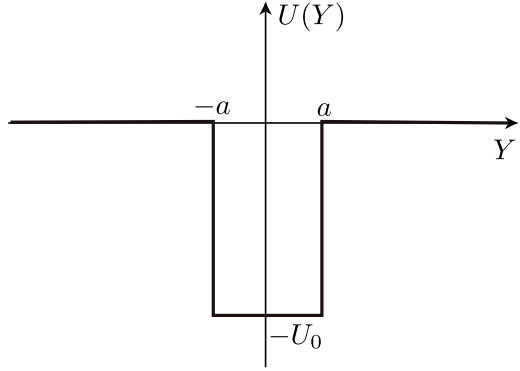}
\end{center}
\caption{The square well potential with a depth $U_0$. There exists a threshold bound state 
if we choose $U_0$ suitably.}
\label{square well potential plot}
\end{figure}

Next solvable example is a square well potential 
\begin{equation}
U(Y)=\left\{
\begin{array}{ll}
-U_0, & |Y|< a\\
0, & |Y| > a
\end{array}
\right..
\label{eq:squr_well_pot}
\end{equation}
(See Fig.\ref{square well potential plot}.)
Then, the coupling function is given by
\[
\e(Y)=\left\{
\begin{array}{ll}
\frac{\sqrt{2U_0}}{m_n}, & |Y|< a\\
0, & |Y| > a
\end{array}
\right..
\]
Using the normalization of the profile $\e(y)$, we determine
\[
U_0 = \frac{m_n^2}{4a g_4^2}.
\]
Energy levels and wave functions for this square well potential can be  exactly solved,
and the threshold bound states occur when we choose 
\begin{equation}
U_0 = \frac{\pi^2}{8a^2} n^2, \qquad n=0,1,2,\ldots.
\label{eq:threshold_squr_well_pot}
\end{equation}
Therefore the mass spectra of threshold bound states are given by
\[
m_n^2 = 4 g_4^2 a U_0 = \frac{\pi^2}{2a}g_4^2 n^2.
\]
Again we find there is a massless zero mode at the level $n=0$
and finite mass gaps for higher excited states.

\section{Four-dimensional Coulomb law 
}\label{sc:coulomb_law}

In this section, we wish to demonstrate that the position-dependent
gauge coupling in Eq.(\ref{eq:gauge_field_lagrangian}) 
exhibits the four-dimensional Coulomb law for 
the static source on the world volume of the domain wall, 
following a treatment in Ref.\citen{Dvali:2000rx}.
We introduce a source term ${\cal J}^a_M$ 
to Eq.(\ref{eq:gauge_field_lagrangian}) to examine the 
response 
\begin{equation}
{\cal L}=-\frac{1}{4}\e(y)F^a_{MN}F^{aMN} + {\cal J}^a_M W^{aM}.
\end{equation}
We are interested in an external source ${\cal J}_M(x, y)$ 
that is localized on the world volume of the domain wall, 
and has components only in four-dimensional world-volume: 
\begin{equation}
{\cal J}^a_M(x, y) =\delta(y) \delta_{M \mu} J^a_\mu (x) . 
\label{eq:source}
\end{equation}

The free field equation for the gauge field $W^a_M$, namely 
the equation of motion ignoring the nonlinear interaction 
terms reads 
\begin{equation}
0=\e(y)(\del^M\del_MW^a_N-\del^M\del_NW^a_M)
+\del^y\e(y)(\del_yW^a_N-\del_NW^a_y)
-{\cal J}^a_N .
\end{equation}
In this section we choose the Lorentz gauge (in five dimensions) 
\begin{equation}
\partial^M W^a_{M} =0 .
\end{equation}
Then the field equation becomes 
\begin{equation}
0= \e(y)
\partial^M \partial_M W^a_N 
+\partial^y\e(y)
(\del_yW^a_N-\del_NW^a_y)-{\cal J}^a_N .
\label{eq:EOM_Wmu_u1}
\end{equation}
Since the source has no $y$-component, 
the field equation for the extra-dimensional component $W_y$ 
becomes 
\begin{equation}
0= \e(y)\partial^M \partial_M W^a_y ,
\label{eq:EOM_Wy_u1}
\end{equation}
without any source term at $y=0$. 
Because of nonnegative definiteness of $\e(y)$, we find that 
$W^a_y$ obeys a free field equation in five dimensions without 
source. 
Assuming that there is no external source at infinity, 
we obtain that there is no nontrivial solution. 
To demonstrate it, let us solve the free field equation 
by going to the Euclidean space. 
Denoting the mixed representation of the momentum space only in 
 four dimensions as $\tilde W^a_y(p, y)$, we obtain 
\begin{equation}
0= (p^2-\partial_y^2) \tilde W^a_y(p, y) .
\end{equation}
There are two independent solutions with $C_+(p), C_-(p)$ as 
arbitrary functions of $p$ 
\begin{equation}
\tilde W^a_y(p, y) =C_+(p) e^{ p y}+C_-(p) e^{- p y}, 
\label{eq:sol_Wy_u1}
\end{equation}
which is valid in the entire region $-\infty < y < \infty$.  
Since no external source at infinity requires 
$\tilde W^a_y(p, y) \to 0$ at both infinities $y\to \pm \infty$, 
we obtain $C_+(p)=C_-(p)=0$, which implies no nontrivial 
solution: $W^a_y=0$.

Taking account of $W^a_y=0$ and Eq.(\ref{eq:source}), 
we obtain the field equation for $W^a_\mu$ as 
\begin{equation}
0= \e(y)
\partial^\nu \partial_\nu W^a_\mu
+\partial^y \left(\e(y)
\partial_y W^a_{\mu}\right)
-\delta(y) J^a_\mu(x) .
\end{equation}

Now we give a simple example to see the Coulomb law on the domain wall.
We assume the weak coupling region (domain wall) is sufficiently thin 
and the coupling profile behaves as
$\e_{\rm thin}(y) = \frac{\delta(y)}{g_4^2}$.

For regularization purposes, we will add 
a large, but finite values for 
the asymptotic bulk gauge coupling 
\begin{equation}
\e_{\rm thin}(y) = \frac{\delta(y)}{g_4^2}+\frac{1}{g_5^2}.
\label{eq:zero_width_limit}
\end{equation}

We consider this simplified situation of 
the zero-width limit of the domain wall in Eq.(\ref{eq:zero_width_limit}), 
and examine the case of the finite width of the domain wall 
later to confirm that our result is unchanged. 
Going again to the Euclidean space, and using the mixed 
representation $\tilde W^a_\mu(p, y)$ of the momentum space 
only in four dimensions, we obtain the field equation as 
\begin{equation}
0= \frac{(p^2-\partial_y^2)}{g_5^2}\tilde W^a_\mu(p, y)
+\frac{\delta(y)p^2\tilde W^a_\mu(p, y)}{g_4^2}
-\frac{\partial_y\left(\delta(y)\partial_y\tilde W^a_\mu(p, y)\right)}{g_4^2}
-\delta(y) \tilde J^a_\mu(p) .
\label{eq:EOM_delta}
\end{equation}
Since the field equation (\ref{eq:EOM_delta}) for $y \not =0$ 
is identical to the $W^a_y$ component, 
we obtain 
\begin{equation}
\tilde W^a_\mu(p, y) =
C_{\mu}^{a+}(p) e^{- p y}\theta(y) 
+C_{\mu}^{a-}(p) e^{ p y}\theta(-y) 
.
\label{eq:sol_bulk}
\end{equation}
Integrating the field equation (\ref{eq:EOM_delta}) in the 
infinitesimal interval between $y =-\ve$ and $y=\ve$, 
we obtain 
\begin{equation}
0= 
-\int_{-\ve}^\ve dy \,
\frac{\partial_y^2\tilde W^a_\mu(p, y)}{g_5^2}
+\frac{p^2}{g_4^2}\tilde W^a_\mu(p, y)
-\frac{1}{g_4^2}\int_{-\ve}^\ve dy \,
\partial_y\left(\delta(y)\partial_y\tilde W^a_\mu(p, y)\right)
- \tilde J^a_\mu(p) .
\label{eq:EOM_delta_y=0}
\end{equation}
By inserting the solution (\ref{eq:sol_bulk}) for $y\not=0$, 
we find that the 
absence of divergent terms such as $(\delta(y))^2$ in 
Eq.(\ref{eq:EOM_delta_y=0}) from the third term 
$\delta(y)\partial_y\tilde W^a_\mu(p, y)$ requires 
\begin{equation}
C_{\mu}^{a+}(p) =C_{\mu}^{a-}(p)\equiv C^a_\mu(p) . 
\label{eq:consistent_cod}
\end{equation}
With this condition (\ref{eq:consistent_cod}), 
the third term vanishes\footnote{
For an arbitrary smooth function $f(y)$, we find 
$\int_{-\infty}^\infty dy \, f \frac{d}{dy}\left(\delta(y)\sgn(y)\right)
=-\int_{-\infty}^\infty \, dy  \frac{df}{dy}\delta(y)\sgn(y)=0$, where
$\sgn(y)\equiv y/|y|$.
One can demonstrate it by using $\delta(y)\equiv \frac{d}{dy}\sgn(y)$ 
and a regularization such as the $\varepsilon \to 0$ limit of 
$\widetilde{\sgn}(y)\equiv \tanh (y/\varepsilon)$, 
or $\widetilde{\sgn}(y)\equiv y/(2\varepsilon)$ for $-\varepsilon<y<\varepsilon$ 
and $\widetilde{\sgn}(y)\equiv \sgn (y)$ otherwise. 
} 
: 
\begin{equation}
-\frac{1}{g_4^2}\int_{-\ve}^\ve dy \,
\partial_y\left(\delta(y)\partial_y\tilde W^a_\mu(p, y)\right) 
=
\frac{pC_\mu(p)}{g_4^2}\int_{-\ve}^\ve dy \,
\partial_y\left(\delta(y)\partial_y\left(\sgn(y)e^{-p|y|}\right)\right) 
=0 . 
\end{equation}
Now the field equation determines the amount of discontinuity 
$2pC^a_\mu(p)$ of the derivative of $\tilde W^a_\mu$ at $y=0$ in 
terms of the source $\tilde J^a_\mu(p)$ leading to 
\begin{equation}
\tilde W^a_{\mu}(p, y=0) =\frac{g_4^2}{p^2+\frac{2g_4^2}{g_5^2}p}
\tilde J^a_{\mu}(p) . 
\label{eq:coulomb_law_abelian_reg}
\end{equation}
In the strong coupling limit $g_5^2\to \infty$ of the bulk asymptotic 
coupling, we finally obtain
\begin{equation}
\lim_{g_5^2\to\infty} \tilde W^a_{\mu}(p, y=0) =\frac{g_4^2}{p^2}
\tilde J^a_{\mu}(p) . 
\label{eq:coulomb_law_u1}
\end{equation}
If we put a static charge as the source 
$\tilde J^a_{\mu}(p) =Q^a \delta_{\mu0}$, we obtain 
the potential in the coordinate space by a Fourier 
transformation of (\ref{eq:coulomb_law_u1}) 
\begin{equation}
 W_\mu^a(x, y=0) 
= \frac{1}{4\pi}\frac{1}{r}g_4^2Q^a \delta_{\mu 0}, 
\end{equation}
with the three-dimensional spatial distance $r$. 
Thus we obtain the Coulomb law in four-dimensional 
world volume as we anticipated.

The intermediate form of our 
potential in Eq.(\ref{eq:coulomb_law_abelian_reg}) 
 turns out to be 
identical to the result in Ref.\citen{Dvali:2000rx}. 
However, let us note two important difference 
of our analysis from that in Ref.\citen{Dvali:2000rx}. 
Firstly, we have introduced the bulk asymptotic coupling $g_5^2$ 
merely as a regularization parameter, and the agreement of 
the potential at an intermediate step is somewhat technical. 
Moreover our starting Lagrangian possesses the 
five-dimensional field strengths $F^a_{MN}$ localized at the 
wall, in contrast to their Lagrangian in Ref.\citen{Dvali:2000rx}
where only four-dimensional 
field strengths $F^a_{\mu\nu}$ are localized. 
This difference results in the presence of the third term 
in Eq.(\ref{eq:EOM_delta}). 
As we noted, this term gives us a consistency condition 
(\ref{eq:consistent_cod}), which does not follow from 
the field equation in Ref.\citen{Dvali:2000rx}. 
Instead they seem to have assumed (quite naturally) 
a symmetry of $\tilde W^a_\mu(p, y)$ under $y\to -y$. 
Thanks to this consistency condition, the third term in 
Eq.(\ref{eq:EOM_delta_y=0}) vanishes and the resultant 
potential has become identical to that in Ref.\citen{Dvali:2000rx}. 

\begin{figure}[t]
\begin{center}
\includegraphics[scale=0.35]{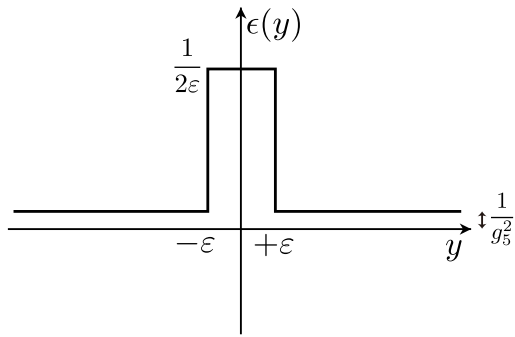}
\end{center}
\caption{The position-dependent coupling is very thin step function. In the limit of $\ve\to0$,
the position-dependent coupling becomes a delta function. An asymptotic height of $1/g_5^2$
is introduced for a regularization.}
\label{step function coupling}
\end{figure}

So far we have been studying 
the zero-width approximation for the domain wall. 
Let us now examine the case of the finite width of the domain wall 
by regularizing the delta function profile. 
As a simplest regularization, we take the following step-function 
ansatz \cite{Dvali:2000rx} 
for the domain wall profile function $\e(y)$ (see Fig.\ref{step function coupling}): 
\begin{equation}
\e_{\rm step}(y) \equiv 
\left\{
\begin{array}{lr}
1/(2\ve) & |y| < \ve \\
1/g_5^2             & |y| > \ve 
\end{array}
\right. 
. 
\label{eq:step_func_reg}
\end{equation}
Since the solution for $W^a_y$ in Eq.(\ref{eq:sol_Wy_u1}) 
uses only the positive definiteness of the coupling 
without referring to the $y$-profile of the coupling 
in Eq.(\ref{eq:EOM_Wy_u1}), we find that $W^a_y =0$ 
even for the finite width case. 
Moreover, the position-dependent gauge coupling 
factors out for $y\not=0$, and the source term exists only 
at $y=0$. 
Therefore the solution for $y\not=0$ as well as 
the discontinuity at $y=0$ are unchanged from 
the zero-width case. 
We thus find that exactly the same solution 
(\ref{eq:coulomb_law_u1}) is valid 
in this finite width case. 
Generally we should obtain somewhat different $y$-profile 
of the solution for other finite width regularizations. 
However, our example shows that 
the qualitative behavior, 
the four-dimensional 
Coulomb law particularly, 
should be the same as 
the zero-width case. 
In the case of Ref.\citen{Dvali:2000rx}, 
the finite width regularization by the step 
function gives a solution for $W^a_\mu$ different from 
the zero-width limit, contrary to our Lagrangian. 
This result arises from the fact that our Lagrangian 
contains the extra-dimensional component of the field 
strength with the position-dependent coupling $\e(y)$.

\section{
Supersymmetric models for BPS walls 
}\label{sc:bps_wall}

\subsection{
General set up 
}\label{sc:setup}
In order to have a realistic brane world with four-dimensional 
world-volume, we construct a domain wall using five-dimensional 
supersymmetric gauge theories, which have 
eight supercharges and consist of vector multiplets and 
hypermultiplets. 
We will consider at least two $U(1)$ vector multiplets labeled by 
$I=1, 2, \cdots$, which contain gauge fields $W^I_M, M=0, 1, \cdots, 4$ 
and neutral scalar fields $\Sigma^I$, besides fermions. 
The vector multiplets for the non-Abelian group $G$ with 
the dimension $\dim G$ also 
contain gauge fields $W^a_M$ and scalar fields $\Sigma^a$ 
in the adjoint representation $a=1, \cdots, \dim G$. 
Hypermultiplets as matter fields contain scalar fields $H_A$ 
(besides fermions) with $A$ labeling different flavors of 
hypermultiplets. 
Non-vanishing values of these $H_A$ will break the $U(1)$ 
gauge symmetries and give domain wall solutions. 
Since we do not wish for non-Abelian gauge symmetry to be 
broken, matter scalars $H_A$ are assumed to be singlets of 
the non-Abelian gauge group. 

It is easy to construct domain walls with a number of 
hypermultiplets interacting with vector multiplets, provided 
the gauge group involves one or more $U(1)$ factors allowing 
the Fayet-Iliopoulos (FI) term.
\cite{Gauntlett:2000ib,Tong:2002hi,Gauntlett:2000bd,
Arai:2002xa,Isozumi:2003rp,Isozumi:2004jc,Isozumi:2004va,
Eto:2004vy,Eto:2005wf,Sakai:2005kz,Eto:2006pg,Eto:2008dm,
Arai:2002ph,Eto:2003ut,{Shifman:2002jm}} 
For simplicity, we use the $U(1)$ gauge groups to build 
domain wall solutions, and the minimal kinetic term for 
$U(1)$ vector multiplets. 
We choose all the FI terms along the common 
direction in the $SU(2)_R$ space. 
We will also use the strong coupling limit of these $U(1)$ 
gauge couplings, whenever we wish to give an explicit exact 
solution of domain walls.  

\subsection{
Domain wall sector 
}\label{sc:wall_sector}

We consider $N_{\rm F} $ charged hypermultiplets%
\footnote{
Although a hypermultiplet contains two complex scalars 
$H^i_A, i=1,2$ for each flavor $A$, we denote only one of 
them $H^1_A$ as $H_A$, 
since the other one $H^2_A$ does not 
participate in our BPS domain wall solutions ($H_A^2=0$). 
} 
$H_A, A=1, \cdots, N_{\rm F}$ with the charge 
$q_I^A, I=1, 2$ for the $U(1)_1\times U(1)_2$ gauge group. 
Neglecting the non-Abelian gauge group in this subsection, 
we obtain the bosonic part of the Lagrangian for the domain wall 
sector 
\begin{equation}
{\cal L}_{\rm wall} = -\frac{1}{4e_I^2}(F_{MN}^I)^2 
+ \frac{1}{2e_I^2}(\partial_M \Sigma^I)^2 
+|{\cal D}_M H
_A|^2 -V,
\end{equation} 
where ${\cal D}_M H
_A=(\partial_M +iW_M^I q^A_I) H_A
$
is the covariant derivative. 
The potential $V$ is given by 
\begin{equation}
V=|(q_I^A \Sigma^I -m_A)H_A|^2 
+\frac{1}{2e_I^2}(Y^{I})^2, 
\end{equation} 
where $m_A \in {\mathbb R}$ is a real mass of the $A$-th 
hypermultiplet. 
The auxiliary fields $Y^{I}$ are given by their 
equations of motion in this model as 
\begin{equation}
Y^{I}
={e_I^2}\left( c_I- q_I^A |H_A|^2\right), 
\end{equation} 
where the FI parameter\footnote{
Both the FI parameter $c^{\alpha I}$ and the auxiliary fields $Y^{\alpha I}$, 
$\alpha=1,2,3$ are triplets of $SU(2)_R$, as in Eq.(\ref{eq:chern_simons_lag}). 
We have chosen the direction of FI parameters for the both 
$U(1)_I, I=1,2$ to be parallel along the third direction 
$c^{\alpha I}=(0,0,c_I)$ and suppress to write components of 
auxiliary fields except the third component which we denote as 
$Y^{\alpha I}=(0,0,Y^I)$. 
} 
for $I$-th $U(1)$ 
factor group is denoted as $c_I$. 

Taking a Bogomolnyi completion, we obtain the Bogomolnyi 
bound which is saturated by the Bogomolnyi-Prasad-Sommerfield 
(BPS) equation. 
The energy density is given by
\begin{eqnarray}
{\cal E}
&=&
\frac{1}{2e_I^2}(\partial_y \Sigma^I -e_I^2(c_I-q^A_I|H_A|^2
))^2 
\nonumber \\
&&+|{\cal D}_y H_A+(q_I^A\Sigma^I-m_A)H_A |^2
\nonumber \\
&&+\partial_y\left(c_I \Sigma^I
-q_I^A \Sigma^I |H_A|^2
+m_A |H_A|^2
\right),  \label{density}
\end{eqnarray} 
where we have chosen the gauge $W_y^I=0$. 
The topological charge for the domain wall connecting the vacuum 
 $\beta$ to $\alpha$ can be read from the last term to give 
the tension 
\begin{eqnarray}
T_{\alpha \leftarrow \beta} 
= \int_{-\infty}^{+\infty} d y \, \partial_y f,  &\text{\ with\ }&
f \equiv c_I \Sigma^I -q_I^A \Sigma^I |H_A|^2 
+m_A |H_A|^2 .\label{Bcompl}
\end{eqnarray} 
The BPS equations are obtained from Eq.~(\ref{density}) as 
\cite{Isozumi:2004jc,Isozumi:2004va,Eto:2004vy,Eto:2005wf,Sakai:2005kz,Eto:2006pg}
\begin{eqnarray}
\left(\partial_y + q_I^A (\Sigma^I + iW_y^I)\right)H_A 
&=& H_A m_A, \label{eq:bps_hyper}
\\
\partial_y \Sigma^I 
&=& e_I^2(c_I - q_I^A |H_A|^2). 
\label{eq:bps_vector}
\end{eqnarray} 
The gauge group indices $I$ are summed for a fixed 
flavor index $A$ in the first BPS equation (\ref{eq:bps_hyper}), 
and vice versa in the second BPS equation (\ref{eq:bps_vector}). 
When these equations are satisfied, the energy density
becomes equal to $[f]^{y=+\infty}_{y=-\infty}$. 

The first BPS equation (\ref{eq:bps_hyper}) can be 
solved in terms of a constant matrix called moduli matrix $H_{0A}$ 
(in our case of $U(1)$ gauge theory, $H_{0A}$ is actually 
a vector) \cite{Eto:2005wf}, \cite{Eto:2006pg}
\begin{eqnarray}
H_A 
= H_{0A} \Omega_1^{-q_1^A/2} \Omega_2^{-q_2^A/2} e^{m_A y}, 
\label{eq:hyper_bps_sol} 
\end{eqnarray} 
where $\Omega_I, I=1, 2$ are given by the solution to the master 
equation \cite{Eto:2005wf}, \cite{Eto:2006pg} 
which gives the solution to other BPS equation 
(\ref{eq:bps_vector}). 
In the strong coupling limit $e_I^2\to \infty$, we can find 
exact and explicit solution $\Omega_I$ from 
the following algebraic equations\cite{Eto:2005wf} 
\begin{eqnarray}
c_I 
= |H_{0A}|^2 \Omega_1^{-q_1^A} \Omega_2^{-q_2^A} e^{2m_A y}. 
\label{eq:master_eq_strong_coupling} 
\end{eqnarray} 

In terms of these $\Omega_I$, 
the hypermultiplet scalars are obtained by 
Eq.(\ref{eq:hyper_bps_sol}), whereas the vector multiplet 
scalars $\Sigma^I$ are given by 
\begin{eqnarray}
\Sigma^I 
= {1 \over 2}\partial_y \log \Omega_I. 
\label{eq:vector_scalar} 
\end{eqnarray}

We first consider the case of four hypermultiplets which allows 
an easy construction of appropriate domain walls, 
and then the case of three hypermultiplets 
as a model with the minimal number of matter fields. 

\subsubsection{
Two copies of two 
charged matter fields 
(four flavor model) 
}\label{sc:4flavors}
The simplest model with a domain wall is the $U(1)$ gauge 
theory containing two charged hypermutiplets with 
different masses.\cite{Arai:2002xa} 
We will just take two copies of such models. 
The charges $q_I^A, I=1, 2$ for gauge group $U(1)_I$ 
and masses $m_A$ of the $A$-th hypermultiplets are given 
in Table.\ref{tb:charge_wall1}. 
\begin{table}[hbt]
\begin{center}
\begin{tabular}{|c|c|c|c|c|}
\hline
 & $H_{A=1}$ & $H_{A=2}$ & $H_{A=3}$ & $H_{A=4}$ \\
\hline
$q_1^A$ for $U(1)_1$ & $1$ & $1$ & $0$ & $0$ \\
$q_2^A$ for $U(1)_2$ & $0$ & $0$ & $1$ & $1$ \\
$m_A$ & ${m \over 2}$ & $-{m \over 2}$ & ${m \over 2}$ & $-{m \over 2}$ \\
\hline
\end{tabular}
\end{center}
\caption{Charge and mass of matter fields (hypermultiplets) 
of the four-flavor model}
\label{tb:charge_wall1}
\end{table}
We also choose FI parameters to be positive 
$c_1=c_2\equiv c >0$. 

The BPS domain wall solution is well-known for the two flavor 
model. 
We take the strong coupling limit $e_I^2\to \infty$ where 
the model reduces to a nonlinear sigma model with the 
$T^*CP^1$ target space, allowing an explicit exact solution 
for a domain wall from 
Eq.(\ref{eq:master_eq_strong_coupling}). 
By choosing the boundary condition for $\Sigma \to \mp m/2$ at 
$y \to \mp \infty$, we find 
\begin{eqnarray}
H_1&=&\sqrt{c} {e^{{m\over 2}(y-y_1)} \over \sqrt{2\cosh m(y-y_1)}}, 
\quad 
H_2=\sqrt{c} {e^{{-m\over 2}(y-y_1)} \over \sqrt{2\cosh m(y-y_1)}}, 
\quad 
\nonumber \\
\Sigma^1 &=&{m\over 2}\tanh m(y-y_1), 
\label{eq:cp1wall1} 
\end{eqnarray} 
where the physical meaning of the moduli parameter $y_1$ is the 
domain wall position. 
Precisely the same form of solution is valid for the second 
copy of the $U(1)$ model, with a moduli $y_2$ for the position of 
another wall 
\begin{eqnarray}
H_3&=&\sqrt{c} {e^{{m\over 2}(y-y_2)} \over \sqrt{2\cosh m(y-y_2)}}, 
\quad 
H_4=\sqrt{c} {e^{{-m\over 2}(y-y_2)} \over \sqrt{2\cosh m(y-y_2)}}, 
\quad 
\nonumber \\
\Sigma^2 &=&{m\over 2}\tanh m(y-y_2). 
\label{eq:cp1wall2} 
\end{eqnarray} 

If we take a difference of these two neutral scalars $\Sigma^I$, 
we find a profile suitable for the position-dependent 
coupling function $\epsilon(y)$, provided $ y_1 < y_2 $ 
\begin{equation}
\Sigma^1-\Sigma^2
={m\over 2}\left(\tanh m(y-y_1)-\tanh m(y-y_2)\right),  
\label{eq:coupl_func_cp1wall} 
\end{equation} 
which is positive definite, and falls off 
exponentially fast towards both infinities
similar to the profile of the position-dependent coupling discussed in \S3. 

\begin{figure}[t]
\begin{center}
\begin{tabular}{ccc}
\includegraphics[scale=0.2]{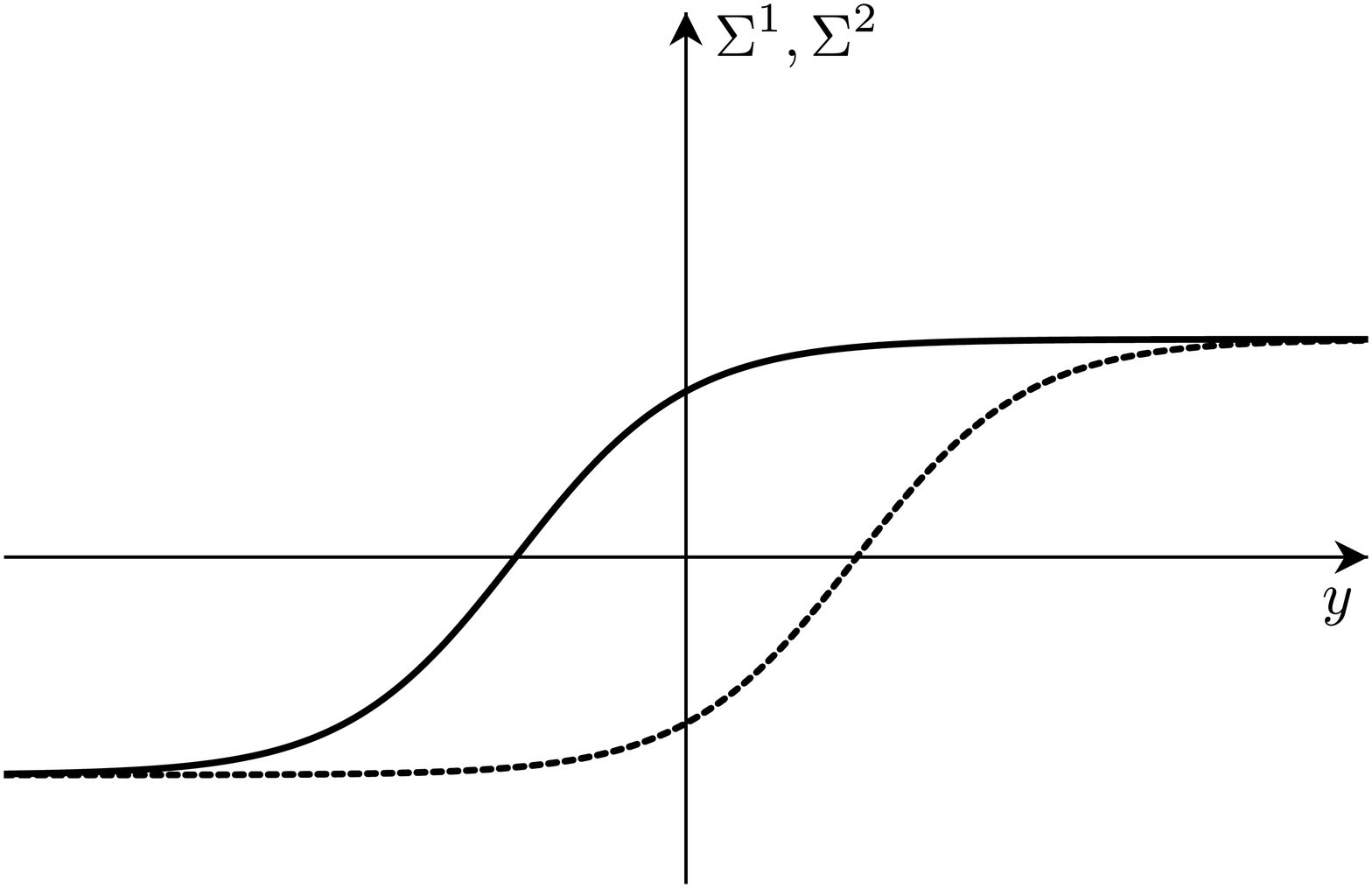} && \includegraphics[scale=0.2]{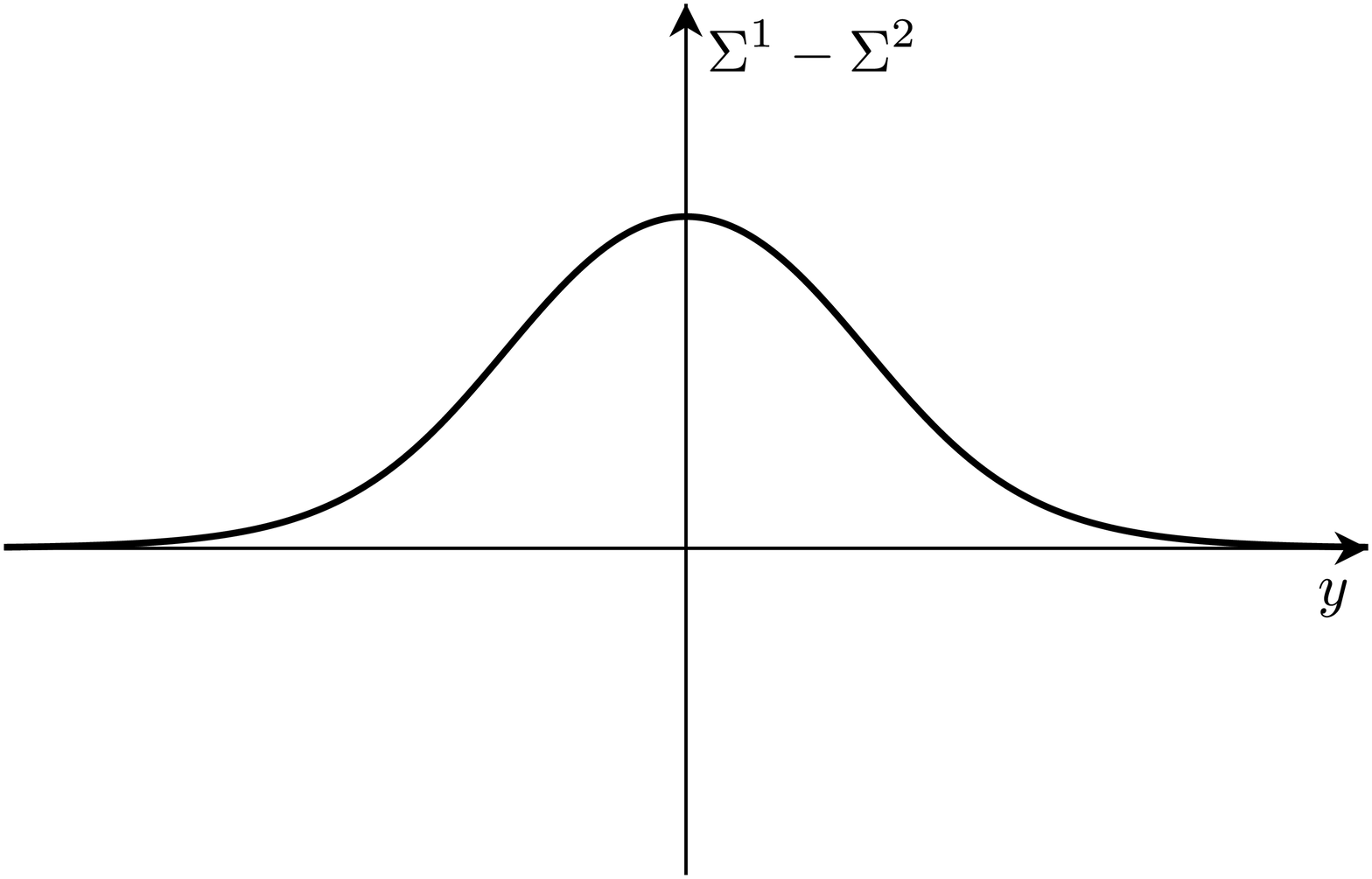}\\
(a) && (b) \\
&\\
\includegraphics[scale=0.2]{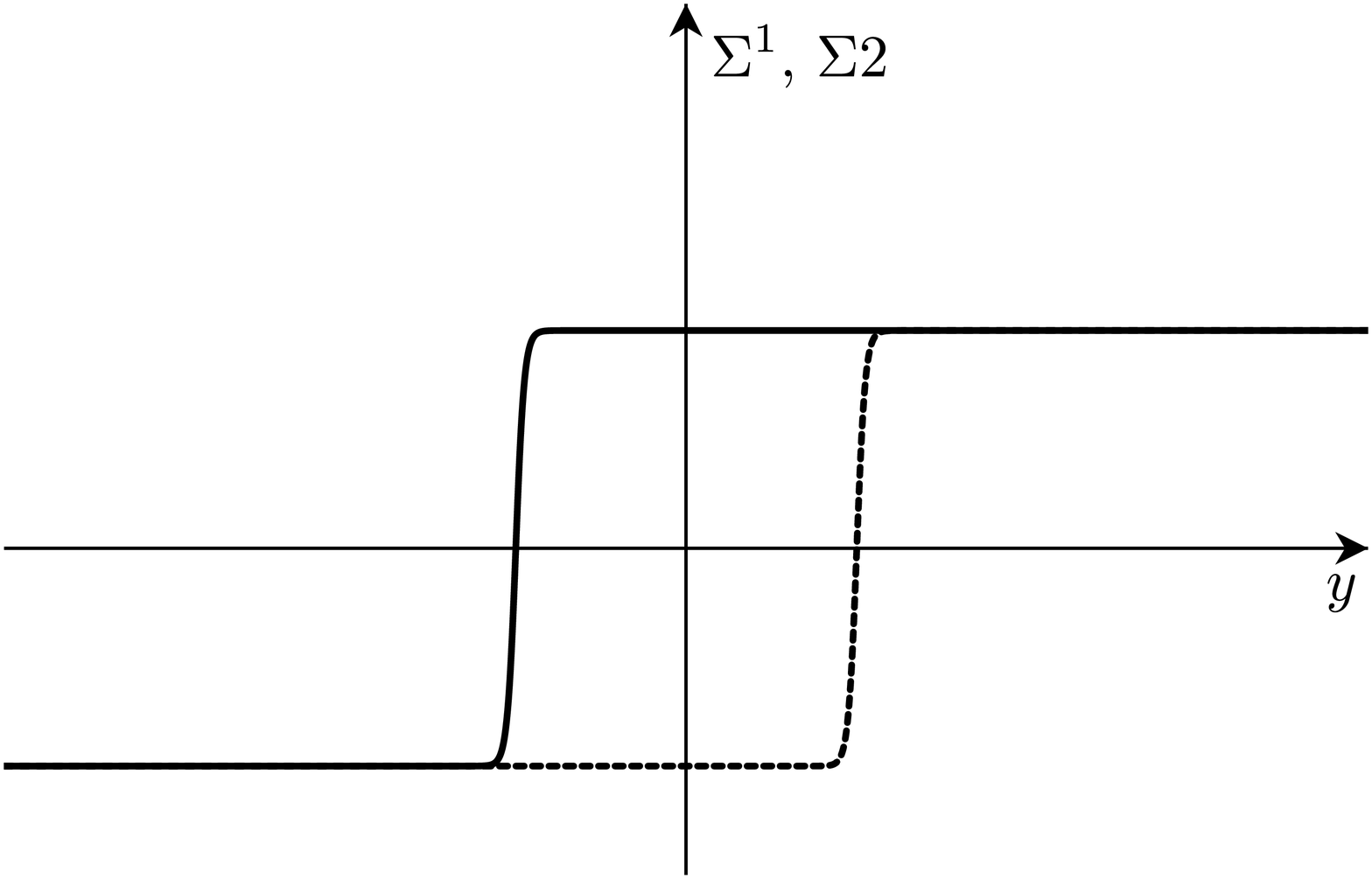} && \includegraphics[scale=0.2]{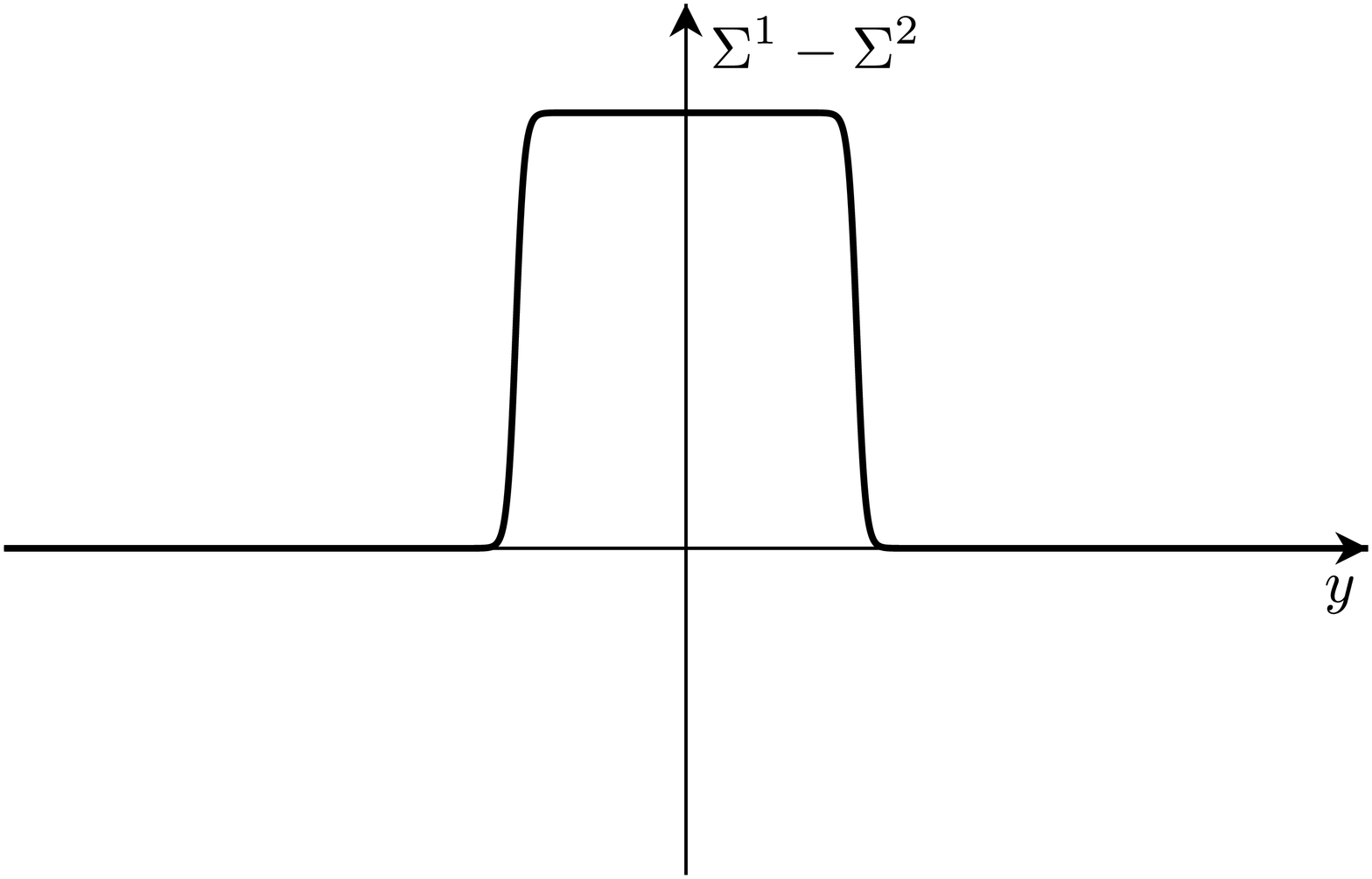}\\
(c) && (d)
\end{tabular}
\end{center}
\caption{A plot of the profile of the domain wall with two Higgs flavors.
The shape of the difference $\Sigma^1-\Sigma^2$ is similar to the position-dependent coupling
discussed in \S3 if $|y_1-y_2|\sim 1/m$. ((a) and (b)) The shape also tends to the step-function
if $|y_1-y_2|\gg 1/m$. ((c) and (d))}
\label{fig:four_flavor}
\end{figure}

The biggest advantage of this model is its simplicity. 
The profile of $\Sigma^1-\Sigma^2$ is positive definite 
for $y_2>y_1$ and has a three layer structure 
with two outer skin with the width $1/m$ and inner wall with 
the width $y_2-y_1$, as shown in Fig.\ref{fig:four_flavor}. 
Therefore we can choose arbitrary wall width by adjusting 
the moduli $y_2, y_1$ whereas the domain wall skin width is fixed 
by the mass parameter of the model $m$. 
The wall profile can be as close as the step function, by 
choosing the mass parameter large $m\to \infty$, with a fixed 
$y_2-y_1$. 
On the other hand, the model becomes unstable for $y_2<y_1$ 
because of negative kinetic term for gauge fields. 
For that reason, one may be tempted to consider an another 
model with no moduli for the adjustable wall width, 
to which we turn next. 

\subsubsection{
$U(1)\times U(1)$ model with three Higgs flavors 
}\label{sc:3flavors}

As is most easily 
seen by taking the strong coupling limit 
(\ref{eq:master_eq_strong_coupling}), each gauge group 
acts as a constraint on hypermultiplet 
scalars to form domain wall solutions. 
Therefore the minimal number of flavors to have domain wall 
solution is the case of $3$ charged hypermultiplets 
$H_A, A=1, 2, 3$. 

Let us take a model containing hypermultiplets with 
the charges $q_I^A, I=1, 2$ and masses $m_A$ as given 
in Table.\ref{tb:charge_wall2}. 
\begin{table}[hbt]
\begin{center}
\begin{tabular}{|c|c|c|c|}
\hline
 & $H_{A=1}$ & $H_{A=2}$ & $H_{A=3}$ \\
\hline
$U(1)_1$ & $1$ & $1$ & $0$ \\
$U(1)_2$ & $0$ & $-1$ & $1$ \\
$m_A$ & $m$ & $0$ & $0$ \\
\hline
\end{tabular}
\end{center}
\caption{Charges and masses of the three matter fields}
\label{tb:charge_wall2}
\end{table}
We easily find two supersymmetric vacua\footnote{
This model has another supersymmetric vacuum, 
which will not be used in this paper: 
the vacuum with $H_1=\sqrt{c_1}, H_3=0$, 
$\Sigma^1=\Sigma^2=m$. 
For the second matter field, the second complex scalar 
of the hypermultiplet $H_2^{i=2}$ has a 
vacuum value $H_2^{i=2}=\sqrt{c_2}$ instead of the first one 
$H_2\equiv H_2^{i=1}=0$, similarly to the model 
considered in Ref.\citen{Eto:2005wf}. 
}. 
The first vacuum is given by 
\begin{equation}
H_1=0, \quad H_2=\sqrt{c_1}, \quad H_3=\sqrt{c_1+c_2}, 
\quad \Sigma^1=\Sigma^2=0.  
\label{eq:1st_vacuum} 
\end{equation} 
The second vacuum is given by 
\begin{equation}
H_1=\sqrt{c_1}, \quad H_2=0, \quad H_3=\sqrt{c_2}, 
\quad \Sigma^1=m, \quad \Sigma^2=0.  
\label{eq:2nd_vacuum} 
\end{equation} 

\begin{figure}[t]
\begin{center}
\begin{tabular}{ccc}
\includegraphics[scale=0.2]{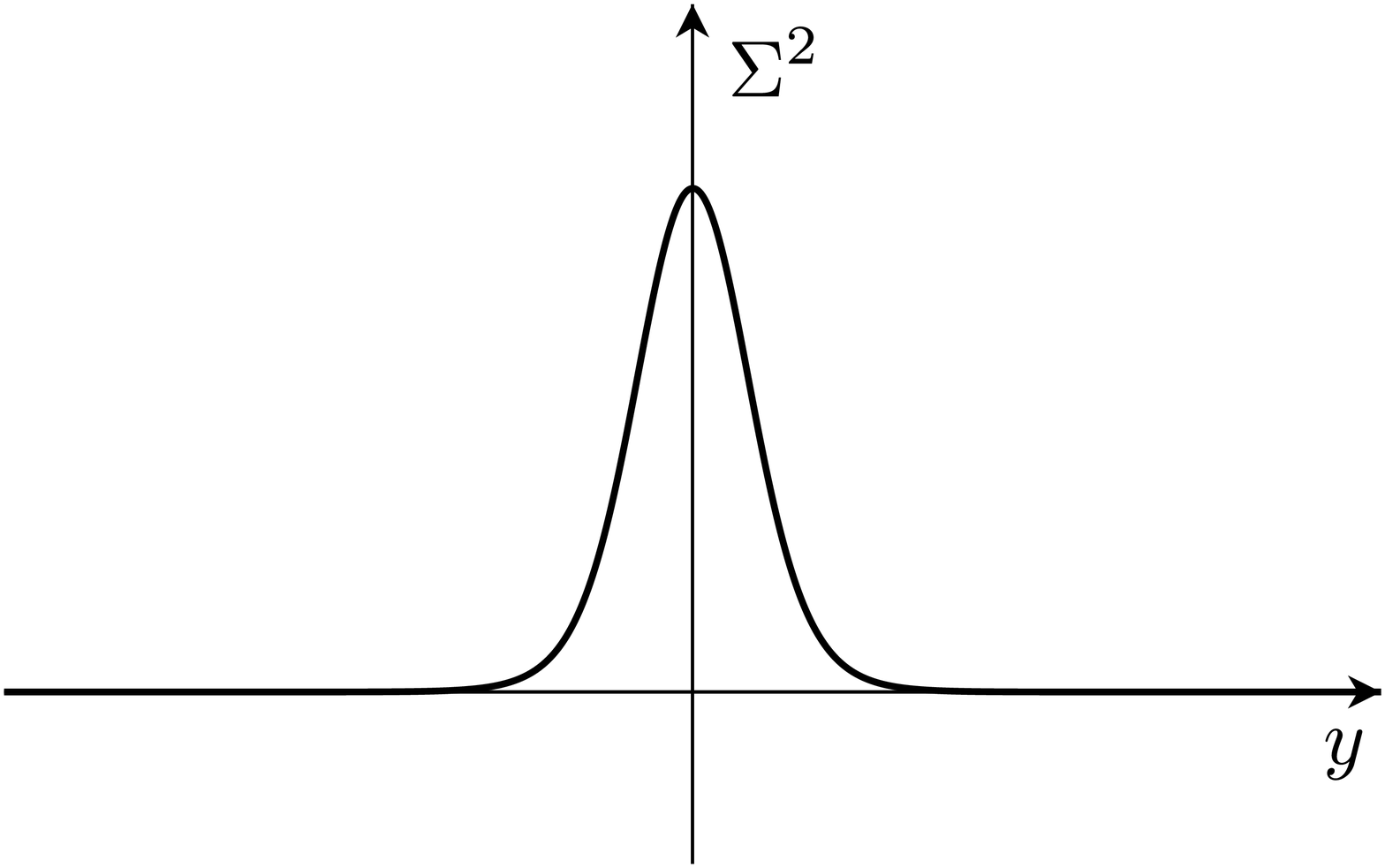}&&
\includegraphics[scale=0.2]{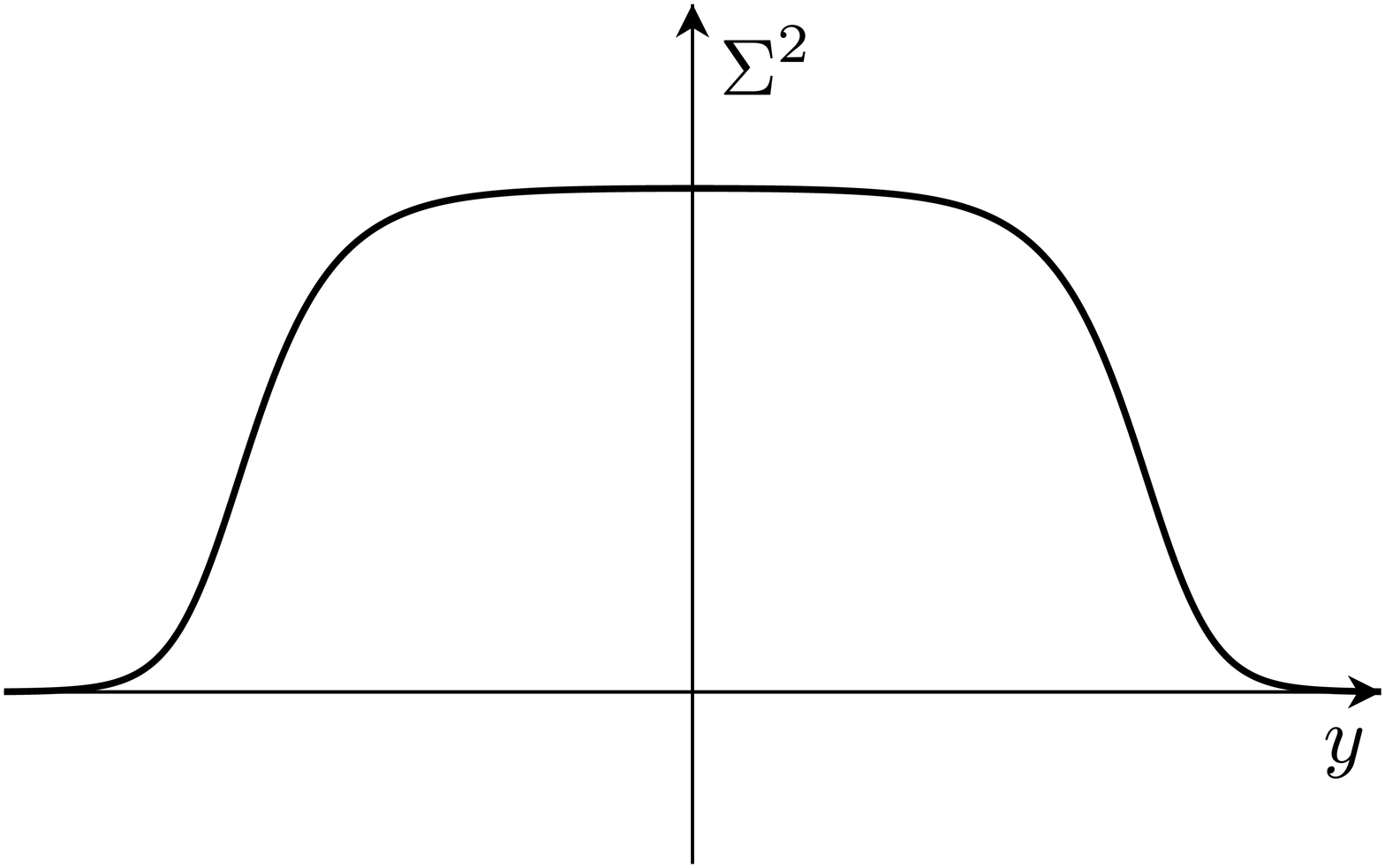}\\
(a) && (b)
\end{tabular}
\end{center}
\caption{A plot of the profile $\Sigma^2$ of the domain wall with three Higgs flavors.
The height of the function is suitably normalized to compare 
the profiles of different FI parameters with each other.
(a) The shape is similar to the position-dependent coupling in \S3 when $c_1/c_2\sim 1$.
(b) We can also obtain the step-function like profile when $c_1/c_2\gg 1$.}
\label{fig:three_higgs}
\end{figure}

Without loss of generality, we can choose the moduli 
matrix to be 
$H_A=(\sqrt{c_1}, \sqrt{c_1}e^{my_0}, \sqrt{c_2})$. 
The moduli parameter $y_0$ is taken to be real \footnote{
The generic moduli $y_0$ is 
complex, whose real 
and imaginary parts correspond to the domain wall position and the 
relative phase of two vacua. 
Since we are not interested in the phase moduli, we take 
$y_0$ to be real here. 
}. 
The BPS domain wall solution connecting these two vacua is 
given in terms of the solution of the master equations 
$\Omega_1, \Omega_2$ as 
\begin{eqnarray}
H_1=\sqrt{c_1} e^{my} \Omega_1^{-1/2}, 
\quad 
H_2=\sqrt{c_1} e^{my_0} \Omega_1^{-1/2} \Omega_1^{1/2}, 
\quad 
H_3=\sqrt{c_2} \Omega_2^{-1/2}, 
\label{eq:hypermultiplet_solution} 
\end{eqnarray} 
where $\Omega_1, \Omega_2$ are given in the strong coupling limit as 
\begin{eqnarray}
\Omega_1 &=& e^{2my} + e^{2my_0}\Omega_2, 
\\
\Omega_2&=&{1-e^{2m(y-y_0)} + 
\sqrt{(1-e^{2m(y-y_0)})^2+4(1+{c_1\over c_2})e^{2m(y-y_0)}} 
\over 2(1+{c_1\over c_2})}. 
\label{eq:solution_master} 
\end{eqnarray} 
The vector multiplet scalars $\Sigma^1, \Sigma^2$ are given by 
(\ref{eq:vector_scalar}). 
Since the domain wall solution connects two vacua in 
(\ref{eq:1st_vacuum}) and (\ref{eq:2nd_vacuum}), 
$\Sigma^2$ is appropriate to give the position-dependent 
coupling function in this model. 
We easily find that the profile of $\Sigma^2$ is reflection 
symmetric with respect to the domain wall position $y=y_0$. 
Let us take the domain wall position at the origin $y_0=0$. 
We obtain 
\begin{eqnarray}
\Omega_2(-y) &=& \left(\Omega_2(y)\right)^{-1}, 
\quad 
\Sigma^2(-y)=\Sigma^2(y). 
\label{eq:symmetry_sigma2} 
\end{eqnarray} 
The asymptotic behavior at large values of $y$ is given by 
\begin{eqnarray}
\Sigma^2(y) \approx m e^{-m(y-{\Delta y\over 2})}, 
\quad 
\Delta y={\log ({c_1\over c_2}) \over m}. 
\label{eq:asymptotic_width} 
\end{eqnarray} 
We see that the domain wall has a three layer structure: 
the overall wall width $\Delta y$ is proportional to $\log (c_1/c_2)$, 
whereas the outer skin has the width $1/m$. 
In this model, both the outer wall (skin) width and 
the inner wall width are fixed by parameters of 
the model, and are not the moduli of the domain wall. 
The exact profile given in Eqs.(\ref{eq:solution_master}) 
and (\ref{eq:vector_scalar}) 
is illustrated for different values of $c_1/c_2$ 
in Fig.\ref{fig:three_higgs}. 

\subsection{
Position-dependent coupling function from the 
cubic prepotential 
}\label{sc:cubic_prepotential}

Interactions between hypermultiplets and vector multiplets 
are specified by charge assignments of the hypermultiplets, 
whereas interactions among vector multiplets are specified by 
the so-called prepotential $a(\Sigma)$. 
It has been found from general principles\cite{Seiberg:1996bd} 
that the prepotential, which gives 
the Chern-Simons coupling for gauge fields together with 
other terms, in our five-dimensional theory 
should be at most cubic in vector multiplets. 
Let us write the bosonic part of a Lagrangian with the 
prepotential $a(\Sigma)$, by denoting the 
group label and gauge generators collectively as $I$. 
Each $U(1)$ factor group can have a triplet of the 
FI term with the parameters $c^{\alpha I}$ 
and the auxiliary fields $Y^{\alpha I}$, $\alpha=1,2,3$. 
Restoring two complex scalar $H^{irA}, i=1,2$ for the 
hypermultiplets with the color ($G$) indices $r$ and the 
flavor indices $A$, we obtain 
\begin{eqnarray}
{\cal L}&=& a_{IJ}\left(-{1\over 4} F_{MN}^I F^{JMN}
+{1\over 2} D_M\Sigma^I D^M \Sigma^J +{1\over 2}Y^{\alpha I}Y^{\alpha J}
\right) -c^{\alpha I} Y^{\alpha I} \nonumber \\
&+&a_{IJK}\left\{-{1\over 24}\epsilon^{LMNPQ}W_L^I
\left(F^J_{MN}F^K_{PQ}
+{1\over 2}[W_M, W_N]^JF^K_{PQ}
+{1\over 16}[W_M, W_N]^J[W_P, W_Q]^K\right)\right\}
\nonumber \\
&+&
 ({\cal D}_M H^{irA})^* {\cal D}^M H^{irA} 
-(H^{irA})^*[(
q^I\Sigma^I-m_A )^2]^r{}_s H^{isA} 
\nonumber \\
&+&
(H^{irA})^*  (\sigma^\alpha)^i{}_j 
q^I(Y^{\alpha I})^r{}_s H^{jsA}
, 
\label{eq:chern_simons_lag}
\end{eqnarray}
where $m_A$ is the mass of the $A$-th hypermultiplet, 
and the derivative of prepotential $a(\Sigma)$ is denoted 
by subscripts like 
\begin{equation}
a_I\equiv {\partial a(\Sigma)\over \partial \Sigma^I}, 
\quad  
a_{IJ}\equiv 
{\partial^2 a(\Sigma)\over \partial \Sigma^I \partial \Sigma^J}, 
\quad  
a_{IJK}\equiv 
{\partial^3 a(\Sigma)\over \partial \Sigma^I \partial \Sigma^J 
\partial \Sigma^K} .
\label{eq:der_prepotential}
\end{equation}
Covariant derivatives ${\cal D}_M$ and $q^I\Sigma^I-m_A$ 
are understood to contain both $U(1)$ 
and non-Abelian components with appropriate charges 
or representation matrices. 
We note that the gauge field kinetic term multiplied by 
the scalar $\Sigma^I$ arises as a supersymmetric completion of 
the Chern-Simons term, both of which follow from 
the cubic prepotential.

The minimal kinetic term for $U(1)$ vector multiplets 
is represented by a term of the form $\Sigma^I\Sigma^I$ in the 
prepotential. 
As given in the previous subsections, the domain wall 
solution leads to a nontrivial kink profile for hypermultiplet 
scalars $H_A^i$, and the vector multiplet scalars $\Sigma^I$. 
Let us call those hypermultiplets and vector multiplets 
participating to form the domain wall as the domain wall sector. 
As described in \S1
, we wish to avoid the bulk in the Higgs phase 
in order to obtain localized gauge fields. 
This requirement is achieved by demanding the domain wall 
hypermultiplets to be neutral under the non-Abelian gauge 
fields which we wish to localize. 
Then the non-Abelian gauge fields can couple to 
the domain wall sector only through the prepotential 
among vector multiplets. 
Since the non-Abelian vector multiplets cannot appear linearly, 
the coupling between non-Abelian vector multiplets and 
domain wall sector should be linear in (a linear combination of) 
$U(1)$ vector multiplets in the domain wall sector. 
By choosing a linear combination $a_1\Sigma^1+a_2\Sigma^2$ 
of two $U(1)$ vector multiplets, we can obtain a 
desired profile of the position-dependent coupling function 
$\epsilon(y)$ at both infinities $y\to \pm \infty$, namely 
the asymptotically vanishing profile with a peak in the middle. 
Therefore we shall consider gauge group to be 
$U(1)\times U(1) \times G$ with $G$ as the non-Abelian 
gauge group which we wish to localize on the domain wall, and 
assume the following prepotential 
\begin{equation}
a\left(\Sigma\right)= {1\over 2e_1^2}(\Sigma^1)^2
+{1\over 2e_2^2}(\Sigma^2)^2 
+{1\over 2}(a_1\Sigma^1+a_2\Sigma^2)\Sigma^a\Sigma^a
\label{eq:prepotential}
\end{equation}
whose constant coefficients $a_1, a_2$ are 
chosen appropriately for each model of the domain 
wall sector. The first and second terms reproduce the 
minimal kinetic terms for the $U(1)$ vector multiplets, as 
we assumed in previous subsections. 

Since the non-Abelian vector multiplet $\Sigma^a$ 
appear only quadratically in the prepotential because of 
gauge invariance, it is easy to see that the above 
prepotential allows the BPS domain wall solution in the 
previous sections to remain a solution to the entire 
system of field equations. 
Therefore we can safely choose the BPS domain wall solution 
as the background solution and consider the effective 
Lagrangian on the domain wall. 
If we choose the following coefficients of the 
prepotential, we obtain the position-dependent gauge 
coupling function $\epsilon(y)$ for the 
non-Abelian gauge fields. 
 For the model with four matter fields 
in \S4.2.1
, we choose 
\begin{equation}
a_1=-a_2\equiv a > 0. 
\label{eq:1stmodel_prepotential}
\end{equation}
 For the model with three matter fields 
in \S4.2.2
, we choose 
\begin{equation}
a_1=0, \qquad a_2 >0. 
\label{eq:1stmodel_prepotential}
\end{equation}
The value of the effective gauge coupling in four-dimensional 
world-volume in Eq.(\ref{eq:profile_function}) can be 
adjusted by choosing the 
value of these coefficients $a$ or $a_2$.

\section{
Conclusion and Discussion 
}\label{sc:discussion}

In this paper, we have discussed the localization of the 
massless vector fields by means of the position-dependent 
gauge coupling. We gave a few concrete examples 
where there exist the massless vector field and mass gap, 
and the Coulomb law emerges inside the domain wall. 
We expect that these localization properties do not depend 
on the details of the concrete coupling functions. 
If the position-dependent coupling function is everywhere 
nonnegative and vanishes at both infinities 
(weak coupling only at the center of the domain wall), 
the localization of the gauge field should occur in a 
similar way.

The position-dependent coupling function desired for the 
localization can be realized by the cubic prepotential 
of the five dimensional supersymmetric gauge theory. 
The coupling function comes about thanks to the profile of 
the domain walls of the Abelian subsectors. 
We expect that these coupling functions offer appropriate 
examples for the localization of non-Abelian gauge fields, 
although the explicit functional forms of the coupling functions 
in our concrete examples of domain wall are somewhat more 
involved than our solvable examples in \S3.

In this paper, we considered the tree-level prepotential, 
in order to obtain the position-dependent coupling. 
The prepotential of the supersymmetric gauge theory with 
eight supercharges receives the nonperturbative quantum 
corrections generally. 
It is an interesting future problem to explore if 
our mechanism of localization of gauge fields due to 
the position-dependent coupling may be realized 
as a result of the nonperturbative quantum effects. 

We have succeeded to localize non-Abelian gauge fields. 
However, we still need to introduce matter fields in 
nontrivial representations of the non-Abelian gauge 
group, in order to build the standard model localized 
on the domain wall. 
It has been found that a non-Abelian flavor symmetry 
for degenerate hypermultiplets provides non-Abelian 
orientational moduli for domain walls.\cite{Eto:2008dm} 
These orientational moduli arises in nontrivial 
representations of the non-Abelian flavor group. 
If we promote (a part of) the flavor symmetry to a local 
gauge symmetry, these non-Abelian orientational 
moduli fields become matter fields interacting 
nontrivially with the non-Abelian gauge fields. 
We can introduce matter fields coupled to the non-Abelian 
gauge fields of our model in this way. 
In order to construct the standard model localized on the 
domain wall, it remains to see if matter fields in 
appropriate (chiral) representations can be introduced 
into our framework. 
The interaction of localized matter fields and the 
localized gauge fields offers an intriguing question of 
charge universality \cite{Dubovsky:2000av}. 
Our model should provide a concrete example of localized 
matter fields assuring the charge universality. 

We have succeeded to embed the BPS domain wall solutions 
in flat space into the five-dimensional supergravity 
theory, and found a model with the warped extra 
dimension.\cite{Arai:2002ph,Eto:2003ut} 
It is also an interesting future problem 
to embed our mechanism of localized 
gauge fields into supergravity. 
In this context, it is worthwhile to examine a recent 
objection against the Fayet-Iliopoulos parameter in 
supergravity.\cite{Komargodski:2009pc}


\section*{Acknowledgements}
One of the authors (N.S.) would like to thank 
David Tong for a collaboration in an early stage, and 
Reijiro Fukuda for a useful discussion on dielectric vacua. 
This work is supported in part by Grant-in-Aid for 
Scientific Research from the Ministry of Education, 
Culture, Sports, Science and Technology, Japan No.21540279 (N.S.),
No.21244036 (N.S.) and No.19740120 (K.O.).

\appendix
\section{Threshold bound states and wave functions} 

In this Appendix, we consider an eigenvalue problem 
of the following static Schr\"odinger equation 
discussed in \S3
\begin{equation}
\left[
-\frac{1}{2}\frac{d^2}{dY^2}+U(Y)
\right]\psi(Y) = E \psi(Y)
\quad \text{for} \quad -\infty < Y < \infty.
\label{appendix:wave equation}
\end{equation}
We will find wave functions of the threshold bound states 
satisfying $E=0$ by choosing a suitable height of the 
potential $U(Y)$. 

\subsection{cosh potential}
Let us first take the potential 
$U(Y)=-U_0/\cosh^2 \alpha Y$ in Eq.(\ref{cosh potential}) 
 (see Fig.\ref{cosh-potential plot}). 
Changing the variable by $\xi = \tanh \alpha Y$ ($-1<\xi<1$) 
and defining $2U_0/\alpha^2 = s(s+1)$, and 
${\cal E}\equiv \sqrt{-2E}/\alpha$, 
(\ref{appendix:wave equation}) becomes 
\begin{equation}
\left[
\frac{d}{d\xi}(1-\xi^2)\frac{d}{d\xi}+s(s+1)-\frac{{\cal E}}{1-\xi^2}
\right]\psi(\xi) =0. 
\label{wave equation in xi}
\end{equation}
In order to find a 
series solution around $\xi=1$, we define 
$u=(1-\xi)/2$ and assume 
$
\psi = u^\gamma \phi(u)=\sum_{l=0}^\infty a_l u^l
$. 
By requiring a 
regular solution at $\xi=1$, 
we find 
$\gamma={\cal E}/{2}$. 
Using $\phi(u)$ instead of $\psi$, 
Eq.(\ref{wave equation in xi}) reduces to 
\begin{equation}
\left[
u(1-u)\frac{d^2}{du^2}
+{1+{\cal E}}(1-2u)\frac{d}{du}
-({\cal E}-s)({\cal E}+s+1)
\right]
\phi(u) =0,
\label{wave equation in u}
\end{equation}
where $\phi(u)\equiv\sum_{l=0}^\infty a_l u^l$. 
In order to obtain solutions regular at $u=1$ ($\xi=-1$), 
the series in $\phi(u)=\sum_la_lu^l$ has to terminate and 
${\cal E}=s-n$. 
Thus the $n$-th eigenfunction is given by the 
following $n$-th order polynomial $\phi_n(u)$ with 
$0\leq n < s$ for a given $s$ 
\[
\phi_n(u)
= a_0\sum_{l=0}^n\frac{1}{l!}\frac{(l+2s-n)!}{(2s-n)!}
\frac{n!}{(n-l)!}\frac{(s-n)!}{(l+s-n)!}(-u)^l.
\]

Threshold bound state is given by ${\cal E}=0$, that is, 
$s=n$, which is achieved by choosing the potential height 
$U_0 = \alpha^2 n(n+1)/2$ as in Eq.(\ref{eq:potential_height}). 
The wave function of the $n$-th threshold bound state is 
given by 
\begin{equation}
\phi_n(u)
=a_0\sum_{l=0}^n\frac{1}{(l!)^2}\frac{(n+l)!}{(n-l)!}(-u)^l.
\label{eq:cosh_threshold_bound_st}
\end{equation}

\subsection{square well potential}

Here we consider the square well potential 
with a finite potential step in Eq.(\ref{eq:squr_well_pot}) 
(see Fig.\ref{square well potential plot}). 
The reflection symmetry $Y\to -Y$ dictates that eigenfunctions 
must be either an even or odd function. 
 For $|Y|<a$, we obtain the eigenfunctions 
$\psi(Y) = A\cos k Y$ for even functions, and 
$B\sin k Y$ for odd functions, 
with $k=\sqrt{2(E+U_0)}$. 
 For $Y>a$, the boundary condition at infinity requires that 
$\psi(Y) = Ce^{-\beta |Y|}$ with $\beta=\sqrt{-2E}$. 
The bound state exists if and only if $E<0$. 
The connection condition at $Y=a$ gives 
$k\tan ka=\beta$ for even functions 
and $k\cot ka = -\beta$ for odd functions. 

The threshold bound state with $E=0$ means that $\beta=0$, 
that is, 
the wave function becomes a constant outside of the well. 
The connection condition for $\beta=0$ is solved by 
discrete values of $k$: $k_n=\frac{n\pi}{2a}$ with nonnegative 
integer $n$, where $n$ is even for the even wave function 
and odd for the odd wave function, respectively. 
We obtain the $n$-th threshold bound state by choosing 
the potential depth $U_0 = k^2/2={n^2\pi^2}/({8a^2})$ 
as in Eq.(\ref{eq:threshold_squr_well_pot}). 
The wave function of the 
$n$-th threshold bound state is given for even $n=2l$
($l=0,1,2,\ldots$)
\begin{equation}
\psi_n(Y) = \left\{
\begin{array}{ll}
A\cos k_n Y & |Y| < a\\
A(-1)^{\frac{n}{2}} &  |Y| > a
\end{array}
\right.,
\end{equation}
and  for odd $n=2l+1$ ($l=0,1,2,\ldots$)
\begin{equation}
\psi_n(Y) = \left\{
\begin{array}{ll}
B\sin k_n Y & |Y| < a\\
B(-1)^{\frac{n-1}{2}} &  |Y| > a
\end{array}
\right..
\end{equation}

%

\end{document}